# *In vivo* stiffness measurement of epidermis, dermis, and hypodermis using broadband Rayleigh-wave optical coherence elastography

Xu Feng,[a,c,1] Guo-Yang Li,[a,c,1] Antoine Ramier,[a,b] Amira M. Eltony,[a,c] and Seok-Hyun Yun,[a,b,c,*]

[a] Harvard Medical School and Wellman Center for Photomedicine, Massachusetts General Hospital, Boston, Massachusetts 02114, USA
[b] Harvard-MIT Division of Health Sciences and Technology, Cambridge, Massachusetts 02139, USA
[c] Department of Dermatology, Massachusetts General Hospital, Boston, Massachusetts 02114, USA
[1] Co-first authors with equal contribution.
* syun@hms.harvard.edu

**Abstract:** Traveling-wave optical coherence elastography (OCE) is a promising technique to measure the stiffness of biological tissues. While OCE has been applied to relatively homogeneous samples, tissues with significantly varying elasticity through depth pose a challenge, requiring depth-resolved measurement with sufficient resolution and accuracy. Here, we develop a broadband Rayleigh-wave OCE technique capable of measuring the elastic moduli of the 3 major skin layers (epidermis, dermis, and hypodermis) reliably by analyzing the dispersion of leaky Rayleigh surface waves over a wide frequency range of 0.1–10 kHz. We show that a previously unexplored, high frequency range of 4–10 kHz is critical to resolve the thin epidermis, while a low frequency range of 0.2–1 kHz is adequate to probe the dermis and deeper hypodermis. We develop a dual bilayer-based inverse model to determine the elastic moduli in all 3 layers and verify its high accuracy with finite element analysis and skin-mimicking phantoms. Finally, the technique is applied to measure the forearm skin of healthy volunteers. The Young's modulus of the epidermis (including the stratum corneum) is measured to be ~ 4 MPa at 4–10 kHz, whereas Young's moduli of the dermis and hypodermis are about 40 and 15 kPa, respectively, at 0.2–1 kHz. Besides dermatologic applications, this method may be useful for the mechanical analysis of various other layered tissues with sub-mm depth resolution.

## 1. Introduction

The skin is a layered tissue characterized by 3 major layers, namely, the epidermis, dermis, and hypodermis. Each layer plays an important role in maintaining the structural integrity and pathophysiological functions of the skin. The epidermis is the outmost protective layer (~ 100 *µ*m in most body sites), consisting of the stratum corneum (SC) and viable epidermis [1, 2]. The dermis is the middle layer with a typical thickness of ~ 1 mm [2], providing nourishment and mechanical support for the epidermis [3]. The hypodermis is the softest layer with a thickness of a few mm. Characterizing the biomechanical properties of individual skin layers *in vivo* is important for understanding the aging process [4], as well as the mechanisms of other dermatological conditions such as scarring [5], skin cancer [6], and inflammatory reactions with edema or fibrosis [7].

Several methods have been developed to measure skin elasticity *in vivo*. Mechanical techniques such as tonometry, indentation, suction, torsion and twisting are useful to measure the various mechanical properties of the skin as a whole, but cannot resolve the individual skin layers [8] and, when applied *in vivo*, are subject to significant variability dependent on experimental conditions [9]. Elastography based on different imaging modalities, such as ultrasound and optical coherence tomography (OCT), have been applied to characterize the layered mechanical properties of the skin. Ultrasound elastography has been used to resolve the Young's modulus in the dermis and hypodermis [10-13]. However, the resolution (~300 $\mu$m) of ultrasound is insufficient to resolve the epidermis. Optical coherence elastography (OCE) takes advantage of the high imaging resolution (~ 10 $\mu$m) of OCT and has been applied to characterizing the skin [14-19]. Compression-based OCE methods measure static strains in the skin layers under a bulk compressional force [6, 19] but cannot directly quantify the Young's moduli of individual skin layers [20].

OCE based on mechanical wave propagation enables quantitative characterization of elasticity with spatial resolution approximately proportional to the wavelength or spatial pulse width of the mechanical wave [21]. Rayleigh surface wave OCE, in particular, offers a promising approach to obtain depth-resolved information as surface waves can be readily excited and measured at the skin's surface without the constraint of imaging depth. For skin tissues, the Rayleigh waves are typically leaky waves with both dissipative and dispersive losses along the surface. The amplitude of surface waves decays exponentially with depth. At any transverse location, the 50%-energy penetration depth is about a half wavelength [22]. Considering the typical thicknesses of human epidermis and dermis and the bulk shear wave velocity of ~ 4 m/s in the dermis, we find that the Rayleigh wave at 3 kHz predominantly occupies the dermis, while the Rayleigh wave at 10 kHz has ~ 50% energy in the thin epidermis. It is evident that high-frequency measurement is essential to resolve the elasticity of individual skin layers.

Several approaches for excitation of Rayleigh waves in skin have been investigated. Air puff excitation, which has a typical bandwidth of a few hundred Hz, has been applied to differentiating macroscopic lesions such as systemic sclerosis [18], but cannot resolve individual skin layers due to the long wavelength (~ 10 mm at 400 Hz). Air-coupled acoustic radiation force excitation has recently been developed to measure skin anisotropy *in vivo* with frequency up to 4 kHz [16]. Zhou et al. [17]

measured Rayleigh waves of frequency < 4 kHz in human skin and developed a weighted-average velocity inversion model to estimate the depth profile of elasticity. However, the lack of sufficient high frequency content obscured features in the dispersion curves, resulting in poor reliability for resolving individual skin layers. In our previous work, we used piezoelectric actuated excitation to generate waves with frequencies up to 16 kHz for Rayleigh-wave OCE in the cornea [23, 24].

Here, we demonstrate broadband OCE measurement (0.1–10 kHz) for *in vivo* quantification of skin elasticity. This study shows, for the first time to our knowledge, the critical benefit of high frequency waves (4–10 kHz) for differentiating the thin epidermal layer from the thicker and softer dermis and hypodermis. The OCE system used has been optimized for the generation and detection of high-frequency waves [23, 24]. We observe feature-rich dispersion curves for human skin and establish a dual bilayer-based model to resolve the distinct stiffnesses of the epidermis, dermis, and hypodermis from the measured dispersion data using simple analytic fitting. We validate this method by finite element analysis (FEA) and tissue phantom experiments and then apply it to human forearm skin *in vivo*.

## 2. The acoustoelastic model of skin tissues

*2.1 A bilayer model and key frequencies*

The frequency dependent penetration of Rayleigh waves in layered structures has been well recognized, and a number of algorithms have been developed to calculate shear or Young's moduli of 2 layers and, in some cases, 3 layers from the frequency-dependent wave velocity (dispersion) curves [17, 25-27]. Access to high-frequency waves (> 4 kHz) should enable inverse calculation for 3 layers with improved global fitting accuracy. However, we have reasoned that a simpler algorithm would be possible considering the differences in the thickness and the bulk wave speed in different skin layers. To explore this, we first consider a bilayer model, as illustrated in Figure. 1(a). This model assumes the layers are elastic, linear, and isotropic. The structure consists of a top layer with a thickness $h$ and Young's modulus $E_1$ (material 1) and a semi-infinite substrate with Young's modulus $E_2$ (material 2). We consider the case of stiff layer on soft substrate, that is, $E_1 > E_2$. Each material supports bulk transverse shear waves with velocity $v_{ti} = \sqrt{E_i/(3\rho_i)}$ where $\rho_i$ is the mass density and subscript $i$ (=1 or 2) denotes material. The derivation of the dispersion relation for surface waves in the bilayer structure can be found in the literature [13], and a summary is provided in Appendix A. The dispersion relation can be described as $v = v(h, E_1, E_2; f)$, where $f$ denotes the wave frequency, and $h, E_1$, and $E_2$ are the three input variables. Other parameters are treated as constants, i.e. Poisson ratio and the density for individual skin layer is assumed to be 0.4999 and 1 g/cm³, respectively.

Figure 1 (b) depicts the general dispersion curve of the Rayleigh surface wave. The phase velocity transitions from a low wave speed at $f \to 0$ to a high wave speed at $f \to +\infty$. In fact, when $f \to 0$ the wave speed $v_2^{(R)}$ is equal to the Rayleigh (R) surface wave speed [28] of the substrate material: $v_2^{(R)} \approx 0.955\, v_{t2}$. The phase velocity then increases slowly to $v_{t2}$ at a critical frequency $f_2$, at which the dispersion curve is divided into a nonleaky branch and a leaky branch [29]. It can be shown that $f_2 \approx \frac{v_{t2}}{6.7\, r_{12}^{1/2}\, h}$, where $r_{12} = E_1/E_2$ is the stiffness ratio of the two materials [29]. Beyond this critical point, the wave becomes a leaky surface wave as its phase velocity is greater than the bulk wave speed $v_{t2}$ of the substrate material, and the leaky wave speed increases rapidly with frequency. When $f \to +\infty$ the phase velocity reaches a plateau at $v_1^{(R)} \approx 0.955\, v_{t1}$, the Rayleigh wave speed of material 1. (Another adiabatic solution exists as the Stoneley interface wave between the two layers, but this wave is ignored as it is not a surface wave.) Approximately, the plateau is reached at a frequency $f_1 \approx \frac{v_{t1}}{h}$. We introduce an intermediate frequency $f_{12} = \sqrt{f_1 f_2}$, at which the velocity is approximately the average of the two bulk shear velocities, $(v_{t1} + v_{t2})/2$. We find $f_{12} = \frac{v_{t2}}{2.6\, h}$, independent of the stiffness ratio.

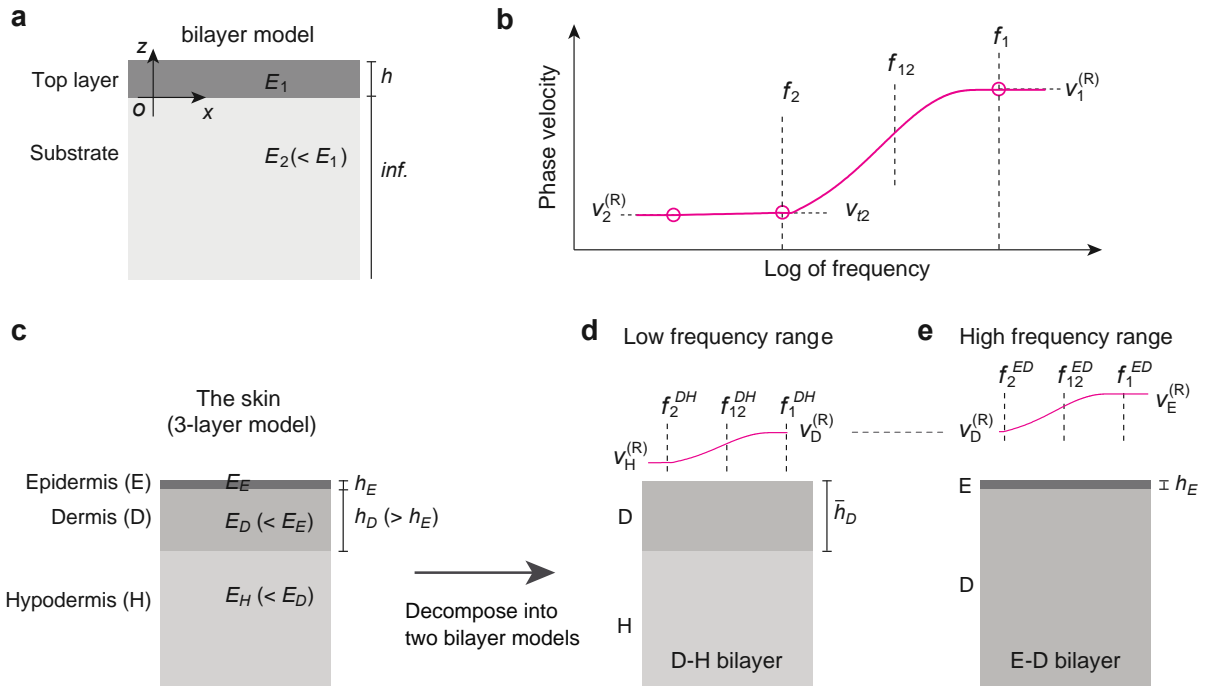

**Fig. 1.** The bilayer models. (a) A basic bilayer model consisting of a top layer with thickness $h$ and a semi-finite substrate. (b) A general dispersion relation of the Rayleigh surface wave in the bilayer model. The dispersion curve is divided into two branches at the transverse wave speed of the substrate ($v_{t2}$). (c) The 3-layer skin model. $h_E$ and $h_D$ are the thickness of the epidermis and dermis, and $E_E, E_D$ and $E_H$ are the Young's moduli of the epidermis, dermis, and hypodermis, respectively. (d) The D-H bilayer model in the low frequency range. $\bar{h}_D$ is an equivalent thickness of the dermis. (e) The E-D bilayer model in the high frequency range. Detailed explanation of the parameters in (d) and (e) can be found in Section 2.2.

*2.2 The skin model and key frequency ranges*

Our main insight is that the 3-layer skin may be decomposed into two bilayer models, the D-H bilayer, as depicted in Fig. 1(d), and the E-D bilayer, as shown in Fig. 1(e), depending on the frequency range. In the lower frequency range, the skin can be modeled as a D-H bilayer: the top layer is the dermis (D), and the hypodermis (H) serves as the substrate. The contribution of the epidermis (E) layer to the Rayleigh surface wave speed can be incorporated by using an equivalent thickness for the dermis, $\bar{h}_D = \bar{h}_D\left(h_D, h_E, \frac{E_E}{E_D}\right)$, where $h_D$ and $E_D$, and $h_E$ and $E_E$ are the thickness and Young's modulus of dermis and epidermis, respectively. We derive $\bar{h}_D$ by considering the equivalent bending stiffness of the epidermis and dermis in the low frequency range. The result is (see Appendix B):

$$\bar{h}_D \approx h_D \left[(1 + \frac{4E_E h_E}{E_D h_D} + \frac{6E_E h_E^2}{E_D h_D^2})/(1 + \frac{E_E h_E}{E_D h_D})\right]^{1/3}. \qquad (1)$$

For typical skin tissues, $\bar{h}_D$ is 1.5 to 1.7 times $h_D$.

In the higher frequency range, the skin can be modeled as an E-D bilayer: the top layer is the epidermis, and the dermis serves as the substrate. In this case, the hypodermis is ignored because the Rayleigh wave with shorter wavelength decays more rapidly with depth, resulting in little penetration into, and hence influence of, the hypodermis.

Table 1 shows the formula to calculate the characteristic frequencies for skin. The distinct lower and higher characteristic frequency ranges corresponding to the D-H and E-D interfaces (respectively) justify our use of the dual bilayer model for skin, with the D-H bilayer used to model skin in the low frequency regime, and the E-D bilayer used in the high frequency regime. As a rule of thumb, an ideal low-frequency range should span from $f_2$ of the D-H layer, $f_2^{DH}$, to the smaller of $f_1$ of the D-H layer and $f_2$ of the E-D layer; that is, $f_2^{DH} < f < min(f_1^{DH}, f_2^{ED})$. Likewise, an ideal high-frequency range should span from the larger of $f_1$ of the D-H layer and $f_2$ of the E-D layer, to $f_1$ of the E-D layer, $f_1^{ED}$. We find that $f_1^{ED}$ is typically a few hundred kHz, a frequency too high to measure with OCE. For reliable curve fitting, we choose the minimum range to occupy up to $0.5 f_{12}^{ED}$, where $f_{12}^{ED}$ is the intermediate frequency of the E-D layer. This minimum high-frequency range can be expressed as $max(f_1^{DH}, f_2^{ED}) < f < 0.5 f_{12}^{ED}$. Table 1 shows the typical values of the characteristic frequencies for two skin-like samples. The estimated low frequency range is approximately 0.1 to 1 kHz, and the high-frequency range is approximately 3 to 10 kHz. To further minimize the impact of hypodermis on E-D bilayer, we use the range of 4–10 kHz in the following study.

**Table 1. Typical values of the characteristic frequencies.** For sample 1: $E_E$=1500 kPa, $E_D$=48 kPa, $E_H$=6.8 kPa, $h_E$=80 μm, $h_D$=1 mm, and $\bar{h}_D$=1.5 mm. For sample 2: $E_E$=4570 kPa, $E_D$=52 kPa, $E_H$=18 kPa, $h_E$=80 μm, $h_D$=1 mm, and $\bar{h}_D$=1.6 mm. $v_{tH}$, $v_{tD}$, and $v_{tE}$ are the bulk shear wave velocities of the hypodermis, dermis, and epidermis, respectively. $r_{DH} = \frac{E_D}{E_H}$, $r_{ED} = \frac{E_E}{E_D}$. The parameters of sample 1 were chosen based on previous literature [13, 17] and are used for the numerical simulation in Fig. 3. The parameters of sample 2 were obtained from the forearm skin of a human (subject #1).

|  | $f_2^{DH}$ | $f_1^{DH}$ | $f_2^{ED}$ | $f_{12}^{ED}$ | $f_1^{ED}$ |
|---|---|---|---|---|---|
| Formula | $\dfrac{v_{tH}}{6.7 r_{DH}^{0.5} \bar{h}_D}$ | $\dfrac{v_{tD}}{\bar{h}_D}$ | $\dfrac{v_{tD}}{6.7 r_{ED}^{0.5} h_E}$ | $\dfrac{v_{tD}}{2.6 h_E}$ | $\dfrac{v_{tE}}{h_E}$ |
| Sample 1 | 0.05 kHz | 2.5 kHz | 1.3 kHz | 19 kHz | 280 kHz |
| Sample 2 | 0.13 kHz | 2.5 kHz | 0.8 kHz | 20 kHz | 490 kHz |

At frequencies between $min(f_1^{DH}, f_2^{ED})$ and $max(f_1^{DH}, f_2^{ED})$, the Rayleigh wave speed is influenced by all three layers. Now we see that for the dual bilayer skin model to be valid, the following two conditions should be met: $f_1^{DH} < 0.5 f_{12}^{ED}$ and $f_2^{DH} < f_2^{ED}$. The first condition requires $\bar{h}_D > 5.2\, h_E$. And the second condition demands $\frac{E_E E_H^2}{E_D^3} < \left(\frac{\bar{h}_D}{h_E}\right)^2$, which reduces to $E_E E_H^2 / E_D^3 < 27$ using the first condition. These two geometrical and stiffness conditions are satisfied for skin tissues.

We can calculate the Young's moduli of all 3 layers using the following approach (Fig. 2):

(1) Determine the Young's modulus of the hypodermis $E_H$. The bulk shear wave velocity of the hypodermis $v_{tH}$ is equal to the measured wave velocity near $f_2^{DH}$ (0.1–0.2 kHz). $E_H = 3\rho v_{tH}^2$, where $\rho = 1$ g/cm³.

(2) Determine the initial estimate of Young's modulus of the dermis $E_{D(0)}$ by fitting the dispersion relation of the D-H bilayer, $v = v(h_D, E_{D(0)}, E_H; f)$ in the low frequency range (0.2–1 kHz); and then determine the initial estimate of Young's modulus of the epidermis $E_{E(0)}$ by fitting the dispersion of the E-D bilayer, $v = v(h_E, E_{E(0)}, E_{D(0)}; f)$ in the high frequency range (4–10 kHz). The epidermal thickness $h_E$ and dermal thickness $h_D$ are determined from the OCT image. If the optical penetration depth of the OCT image is smaller than $h_D$, then set $h_D = 1$ mm. $E_{D(0)}$ is an overestimation of $E_D$, and $E_{E(0)}$ is an underestimation to $E_E$ by ~ 10%.

(3) Calculate the equivalent thickness of the dermis $\bar{h}_D = \bar{h}_D(h_D, h_E, \frac{E_{E(0)}}{E_{D(0)}})$ (Eq. (1) or Eq. (B.3) in Appendix B). ($\bar{h}_D$ is not so sensitive to $\frac{E_{E(0)}}{E_{D(0)}}$. The bias in $\bar{h}_D$ propagating from $\frac{E_{E(0)}}{E_{D(0)}}$ is ~ 2%.)

(4) Determine $E_D$ by fitting the dispersion relation of the D-H bilayer again, $v = v(\bar{h}_D, E_D, E_H; f)$ in the low frequency range (0.2–1 kHz), and then determine $E_E$ by fitting the dispersion of the E-D bilayer again, $v = v(h_E, E_E, E_D; f)$ in the high frequency range (4–10 kHz).

(5) Output $E_E$, $E_D$, and $E_H$.

In step (2) and step (3), the curve fitting uses root-mean-square-error as an objective function. Best fitting is found by searching for the minimum error. Because there is only one variable in each fitting, the fitting complexity is greatly reduced compared to multi-parameter fitting used in inversion models.

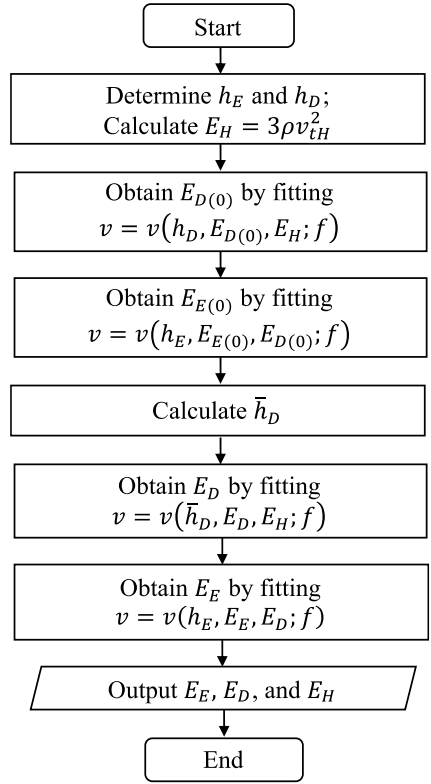

**Fig. 2**. Flowchart of the inverse approach to determine the Young's modulus of the three layers.

*2.3 Verification with numerical simulation*

We used finite element analysis (FEA) to validate the theoretical dispersion relation of surface wave propagation in skin and our inverse approach (details on the FEA can be found in Supplementary Information S1). We built the FEA model using commercial software (Abaqus 6.12, Dassault Systèmes). The model contains three layers, each with different thickness and Young's modulus, to simulate the epidermis ($h_E = 80$ μm, $E_E = 1500$ kPa), dermis ($h_D = 1$ mm, $E_D = 48$ kPa), and hypodermis (semi-infinite, $E_H = 6.75$ kPa). The density of each layer is assumed to be constant, $\rho = 1$ g/cm³. Surface waves are excited by applying local oscillating pressure on the surface. Figure 3(a) shows the simulated cross-sectional displacement profiles of the surface waves at different frequencies. At 0.25 kHz, the wavelength of the surface wave is much greater than the combined thickness of the epidermis and dermis. At 4 kHz, the surface wave is leaky in the hypodermis and much of its elastic energy resides in the dermis. At 10 kHz, the surface wave is leaky in both dermis and hypodermis, and a considerable part of its elastic energy is present in the epidermis.

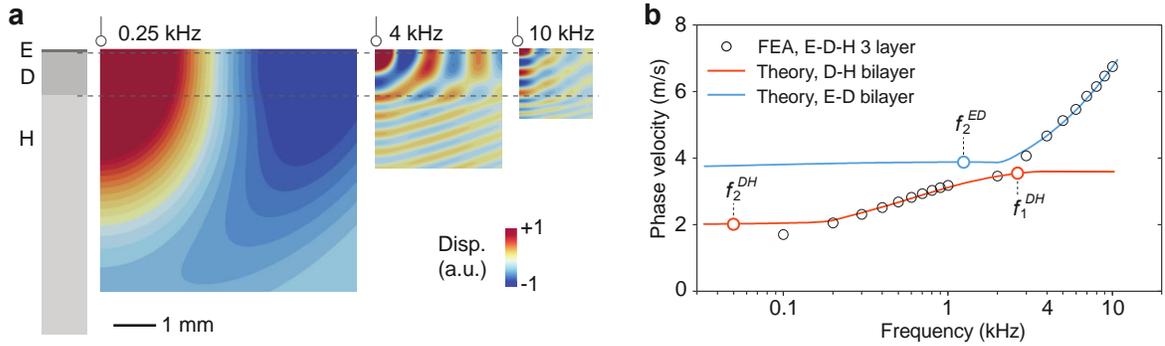

**Fig. 3.** Numerical simulation of surface waves in skin. (a) Finite element simulations showing the vertical displacement profiles of leaky Rayleigh surface waves in the 3-layer model at three representative frequencies: 0.25 kHz, 4 kHz, and 10 kHz. The location of the excitation sources for each plot are indicated at the top left by a circle-and-line). (b) Dispersion relation of the Rayleigh surface waves. Black circles, FEA simulation data. Red curve: Analytical theory obtained from the D-H bilayer model. Blue curve: Analytical theory obtained from the E-D bilayer model.

Figure 3(b) shows the phase velocities (black circles) obtained by FEA at various frequencies in the range of 0.1 to 10 kHz. By performing the inverse analysis on the FEA data we obtain the Young's modulus for each layer. The Young's modulus of the hypodermis inferred from the phase velocity at 100 Hz is 8.7 kPa, which is ~28% larger than the input value. This overestimation is due to $f_2^{DH} \approx 50$ Hz < 100 Hz (see Table 1), which can be eliminated if the phase velocity below 50 Hz is available. Such a bias in the Young's modulus of hypodermis has a negligible effect on subsequent analysis. We further estimate the Young's moduli for the dermis and epidermis as 47 kPa and 1530 kPa, which are in excellent agreement with the values we used in the numerical simulation (48 kPa (-2%) and 1500 kPa (+2%)). The final fitting curves are shown in Fig. 3(b). The key frequencies, $f_2^{DH}$, $f_1^{DH}$, and $f_2^{ED}$, described in the Section 2.2 and Table 1 are also indicated (red and blue circles).

## 3. Experimental validation using skin-mimicking 3-layer phantoms

*3.1 OCE system*

The OCE system has been described previously [23, 24]. In brief, the system uses a swept-source laser with a center wavelength of 1300 nm and a bandwidth of 108 nm at a sweep rate of 43 kHz. The axial resolution is ~ 15 µm. In the sample arm, the laser beam is scanned by a pair of galvanometer mirror scanners (Cambridge Technology, 6210H), and focused by a wide-aperture scan lens (Thorlabs, LSM54-1310) yielding a long working distance of 64 mm and a transverse resolution of ~ 30 μm. The illumination power on the skin is ~ 15 mW, which complies with the ANSI-Z136.1-2014 safety standard. An input/output board (National Instruments, USB-6353) is used to generate analog waveforms for the galvanometer scanners and a piezoelectric actuator used to excite surface waves. The wavelength sweep cycle of the laser is used to synchronize data acquisition, beam scanning, and probe actuation, providing an absolute phase reference for the detection of surface waves.

Mechanical stimulation was achieved using a custom-made contact mechanical actuator. The tip of the actuator (Fig. 4a) is 3D-printed from a biocompatible polymer (Formlabs Surgical Guide Resin). It has the shape of a triangular prism, of which a 2 mm-long edge contacts the skin while the opposing rectangular face is glued to a piezoelectric transducer (PZT) (Thorlabs, PA4CEW). For this study, pure tone stimuli were chosen with a frequency range of 0.1–10 kHz. The exact frequencies tested range from 120 Hz to 840 Hz with an interval of 120 Hz and then from 1 kHz to 10 kHz with an interval of 1 kHz. Above 10 kHz, the increasingly short wavelength and low displacement amplitude made OCE measurements unreliable.

The system is operated in M-B mode. At each transverse location, $m$ consecutive A-lines were acquired. After completing an M-scan, the sample arm beam was moved to the next transverse location and the measurement was repeated. In total, 96 transverse positions were scanned. At each transverse location, the acquisition time was about 0.4 s with $m$ = 172 for each of the 10 stimulus frequencies between 1 and 10 kHz, and the acquisition time was 1 to 5 s with $m$ = 400 for each of the 7 stimulus frequencies between 120 and 840 Hz. In total, an M-scan at 17 frequencies took about 20 s. Since the propagation length decreases with increasing frequency, the sample arm scan length $L$ was varied inverse-linearly with frequency: $L$ [mm] = 8.5 – 0.5*$f$ [kHz] at 1 to 10 kHz, $L$ = 8 mm for 0.24 to 1 kHz, and $L$ = 14 mm for 0.12 kHz.

The propagation of the Rayleigh surface wave was analyzed from the displacement measured at the surface using the method previously described [23, 24]. In summary, we extracted displacement profiles over time $t$ at each transverse location, and then performed a 1-dimensional Fourier transform to move the data from time $t$ domain to frequency $f$ domain. The frequency domain data was then filtered at the driving frequency to obtain lower noise waveforms. After we obtained the displacement profiles over the $x$ coordinate, another 1-dimensional Fourier transform moved the data from the spatial $x$ domain to the wavenumber $k_x$ domain. The wavenumber $k$ of the surface wave was then determined from the plot by selecting the peak corresponding to the Rayleigh surface wave. This filtering in the $k_x$ domain is critical to remove other higher-order modes especially at high frequencies [30]. The phase velocity is then $v = 2\pi f/k$.

*3.2 Phantom validation results*

To validate the method we prepared 3-layer skin-mimicking phantoms [15, 31]. First, a hypodermal layer was prepared using a hydrogel with 3% gelatin concentration on a 35-mm Petri dish. Then, a gelatin hydrogel layer of 7% concentration was deposited for the dermis. Finally, a thin polydimethylsiloxane (PDMS) layer for the epidermis was deposited using a standard 10:1 mixing ratio of base elastomer and curing agent (Sylgard 184, Dow Corning). Two different phantoms with slightly different layer thicknesses were investigated in this study, as shown in Table 2. The thicknesses of the first and second layers were measured using OCT assuming a mean refractive index of 1.4. The total thickness of each phantom was ~ 10 mm.

The measured reference material properties are shown in Table 2. Details for the measurement can be found in Supplementary S2. In brief, the shear wave velocities of the middle and bottom layer were obtained from two bulk phantoms with 7% and 3% gelatin concentrations prepared from the same batch. The shear wave velocity of the top layer was measured from the 150 μm PDMS film detached from phantom #2. The shear wave velocity was estimated by fitting the dispersion curve with a Lamb wave model [24].

**Table 2.** Specifications of the tissue phantoms and measured data.

|  |  | Phantom #1 | Phantom #2 |
|---|---|---|---|
|  | Top | 80 μm | 150 μm |
| Layer thickness | Middle | 0.82 mm | 1.15 mm |
|  | Bottom | 9 mm | 9 mm |
| Young's modulus, reference (mean ± std.) | Top | 1.13 ± 0.08 MPa | |
|  | Middle | 27 ± 2 kPa | |
|  | Bottom | 9 ± 1 kPa | |
| Young's modulus, estimated by OCE (mean ± std.) | Top | 1.14 ± 0.03 MPa | 1.12 ± 0.05 MPa |
|  | Middle | 23 ± 0 kPa | 26 ± 2 kPa |
|  | Bottom | 9 ± 0 kPa | 10 ± 2 kPa |

Figure 4(b) shows the setup of the piezoelectric excitation source and sample. Figure 4(c) shows a representative intensity image, in which the three layers are readily distinguished. The cross-sectional vibrography images reveal the vertical displacement profiles of the surface wave excited at different frequencies (Fig. 4(d)). As the frequency increases, the surface wavelength and propagation length, as well as the depth penetration of the surface wave decrease. Note that the full penetration depth of the surface waves at low frequencies such as 0.48 and 1 kHz is not visible on the images because of insufficient optical SNR below a depth of approximately 1 mm.

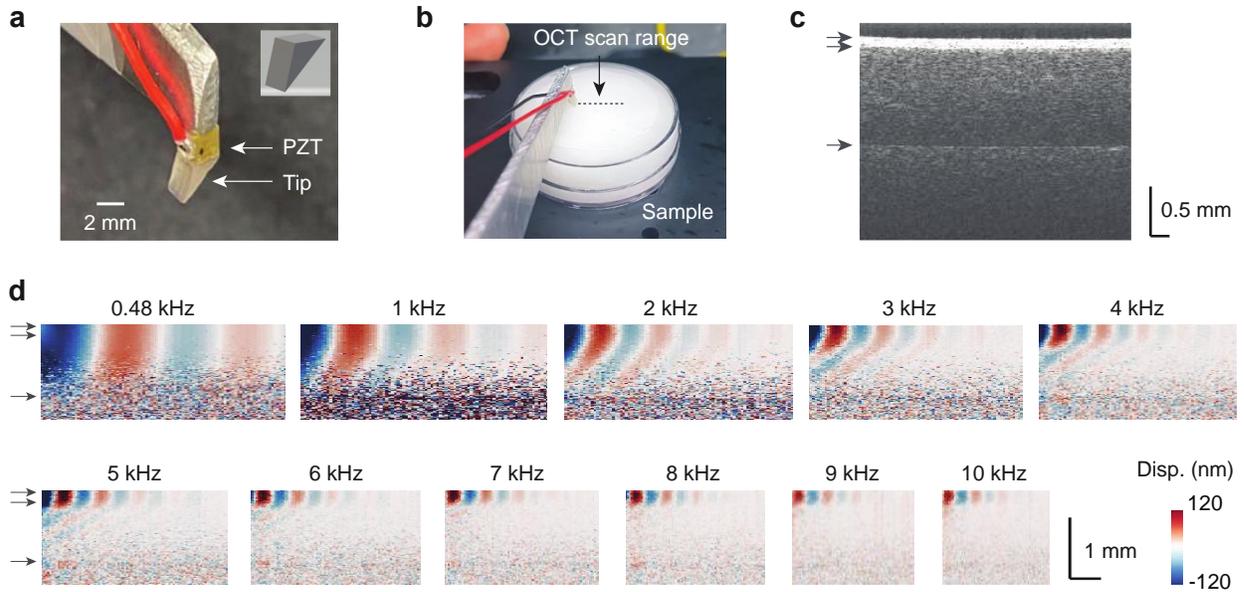

**Fig. 4.** Surface wave propagation in a three-layer skin-mimicking sample. (a) Picture of the contact actuator composed of a piezoelectric transducer (PZT) and a prism-shaped plastic tip. The inset is a three-dimensional drawing of the tip. (b) Picture of the 3-layer phantom. Dashed line, the OCT beam scan path. (c) OCT image of the phantom. Arrows indicate the demarcation lines between layers. (d) Displacement profiles of surface waves excited at different frequencies.

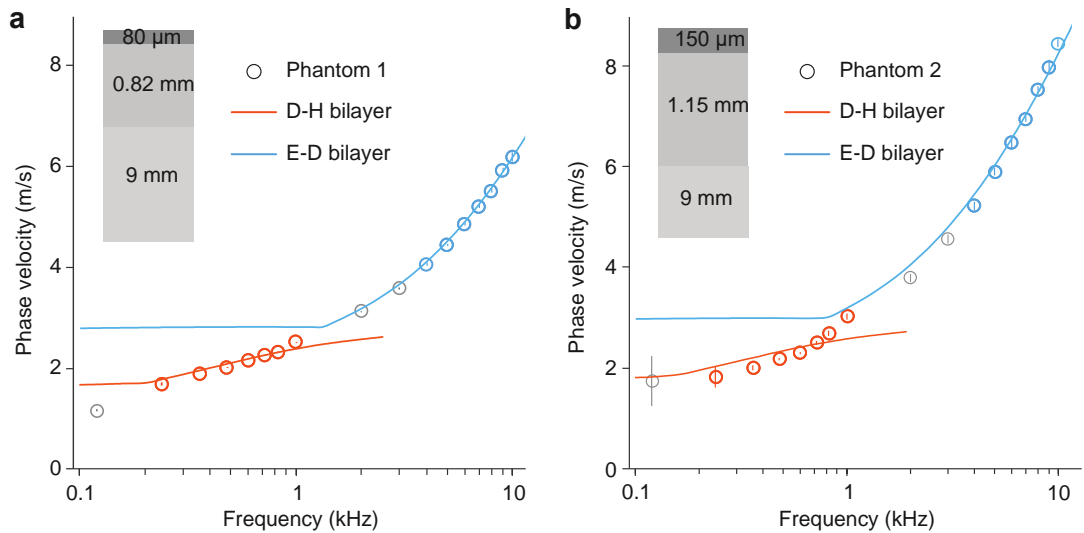

**Fig. 5.** Phase velocity dispersion curve and model fitting in (a) phantom #1, and (b) phantom #2. Dots and error bars, mean and standard deviation of three measurements at three different locations. Red line, D-H bilayer model with best fit in 0.24 to 1 kHz. Blue line, E-D bilayer model with best fit in 4 to 10 kHz.

Figure 5 shows the phase velocity dispersion curves measured from the two phantoms and corresponding fitting results. The wave speed measured at the lowest frequency 0.12 kHz was often erroneous because the wavelength of the wave (~ 20 mm) is greater than the thickness of the sample (10 mm), so the wave speed measurement is influenced by the spurious waves reflected from the bottom interface between the 3% gel and the plastic dish. Therefore, the phase velocity measured at 0.24 kHz was used for estimating the Young's modulus of the bottom 3% gel layer. Table 2 shows the estimated Young's modulus matches well with the reference values.

## 4. Results on human skin *in vivo*

We performed *in vivo* measurements of the dorsal forearm skin of two healthy volunteers (Subject #1: male; Subject #2: female, both in early 30s of age). The study was conducted at the Massachusetts General Hospital (MGH) following approval from the Institutional Review Board (IRB) of Massachusetts General Hospital and the Mass General Brigham Human Research Office. Written informed consent was obtained from both subjects prior to the measurement and all measurements were performed in accordance with the principles of the Declaration of Helsinki. Hairs in the measurement area in the forearm were gently removed using an eyebrow razor. Figure 6 presents the measurement results from Subject #1. The epidermis/dermis interface was segmented using a custom algorithm by analyzing light intensity in the OCT image (Fig. 6(b)). The thickness of the epidermis was calculated as the average optical thickness between the two interfaces divided by an assumed refractive index of 1.4. The dermis/hypodermis interface was not clearly identified due to the limited optical penetration depth of the OCT system. The typical dermal thickness of healthy forearm skin has been reported to be 1.08 ± 0.16 mm (mean ± std.) [2]; therefore, the thickness of the dermis was assumed to be 1 mm for both subjects. Figure 6(c) shows the cross-sectional displacement profiles at different frequencies.

Figure 7 shows the displacement waveforms and the corresponding wavenumber domain plots at different frequencies. Besides the clearly defined Rayleigh wave peak, the secondary peak corresponding to a higher-order fast elastic wave appearing at high frequencies > 4 kHz. This fast wave is the supershear surface wave, which is excited more efficiently by the piezoelectric actuator as its wavelength becomes < 20 mm [30]. The Rayleigh wave velocity was determined from the primary peak in the wavenumber profiles.

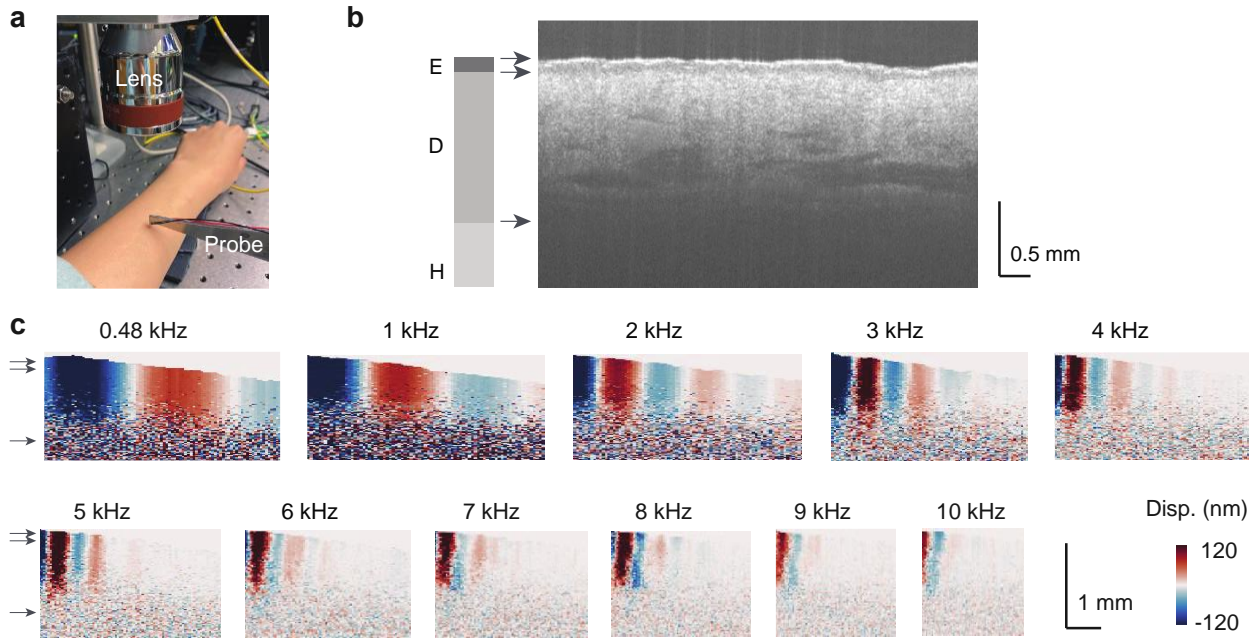

**Fig. 6**. Surface wave propagation in human skin *in vivo*. (a) Picture of a forearm during measurement. (b) OCT images of forearm (E: epidermis, D: dermis, H: hypodermis). Arrows indicate the demarcation lines between layers. (c) Displacement profiles of surface waves excited at different frequencies. The skin layers are labeled on the 0.48 kHz and 5 kHz images.

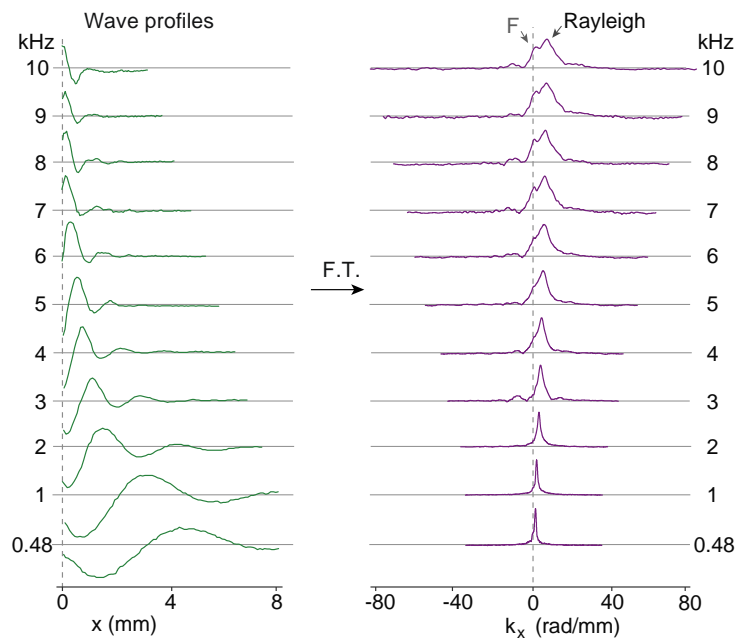

**Fig. 7.** Extracting wave speeds from skin. Left: displacement waveforms as a function of the propagation distance. Right: the corresponding spatial frequency representation obtained by the Fourier Transform. In the spectra at higher frequencies (> 4 kHz), the primary peak with a higher wave number comes from the Rayleigh surface wave, whereas the secondary peak with a lower wave number is due to fast (F) supershear elastic wave.

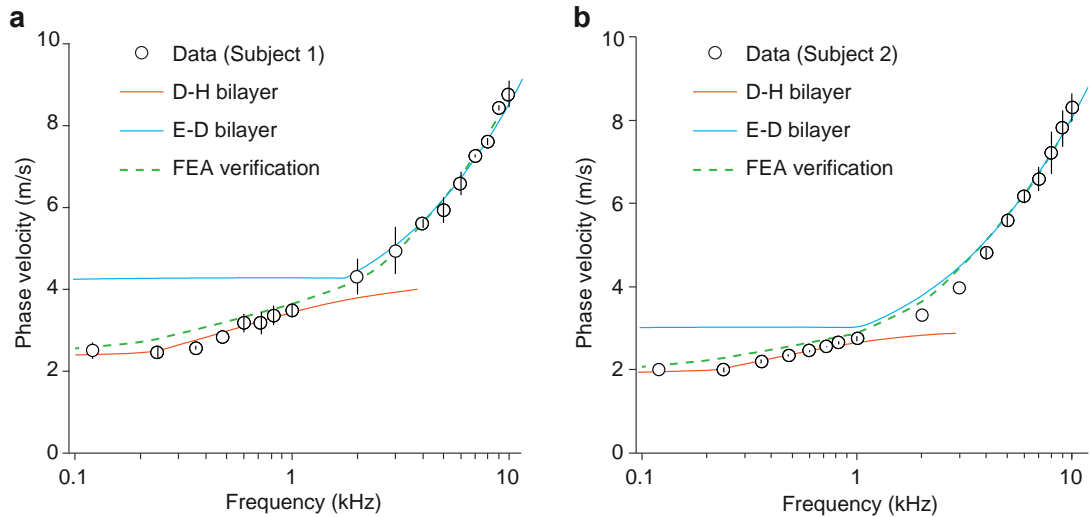

**Fig. 8.** Phase velocity dispersion curve and model fitting for (a) Subject #1, and (b) Subject #2. Dots and error bars, mean and standard deviation of three measurements at approximately the same location. Red line, D-H bilayer model with best fit in 0.24–1 kHz. Blue line, E-D bilayer model with best fit in 4–10 kHz. Green dashed line, FEA verification using the parameters determined by the dual-bilayer algorithm.

Figure 8 compares the measured phase velocity dispersion curves of the two human subjects and the corresponding fitting results. Each data point and its error bars correspond to the mean and standard deviation of three OCE scans that were performed at the same location. Between measurements, the excitation actuator was displaced from the skin and then brought back to approximately the same location on the tissue with the same gentle contact pressure. The repeatability of the measurement was limited by subject motion. The wave velocities measured at 0.12 kHz were slightly higher than those at 0.24 kHz. This could possibly be explained by the interference of spurious waves reflected from the boundaries, such as the muscle and bone underneath the hypodermis. For both subjects, fitting results demonstrate clear transitions of the experimental data from the D-H bilayer in the low frequency regime to the E-D bilayer in the high frequency regime. Table 3 displays the measured Young's modulus for the two subjects. The elasticity values vary significantly between the two subjects in all three layers.

To validate our inverse approach, we performed additional finite element analyses with the parameters in Table 3 and plotted the theoretical dispersion relations of the Rayleigh surface waves. The simulations show good agreements with the experiments (Fig. 8 and Supplementary Fig. S3).

**Table 3.** Young's moduli in the forearm skin tissues of two human subjects

|  |  | Subject #1 | Subject #2 |
|---|---|---|---|
| Layer thickness | Epidermis | 80 µm | 82 µm |
|  | Dermis | 1 mm | 1 mm |
|  | Hypodermis | - | - |
| Measured Young's modulus (mean ± std.) | Epidermis | 4.57 ± 0.27 MPa | 3.48 ± 0.14 MPa |
|  | Dermis | 52 ± 7 kPa | 27 ± 1 kPa |
|  | Hypodermis | 18 ± 2 kPa | 12 ± 1 kPa |

## 5. Discussion and conclusion

Using an advanced OCE system optimized for efficient excitation and detection of surface waves, we have measured wave phase velocities in human skin *in vivo* over a wide frequency range up to 10 kHz, allowing us to accurately quantify the Young's modulus of the different tissue layers. We have shown that high frequency measurement is essential to resolve the thin epidermis, while low frequency measurement permits access to the deep hypodermis. Previous studies have been restricted to frequencies < 4 kHz because of limited sensitivity to the reduced displacement amplitude at higher frequencies. For the same wave energy, the displacement amplitude of an elastic wave decreases as the square of the frequency. That is, the displacement amplitude at 10 kHz would be ~ 6.3 times smaller than the amplitude at 4 kHz. To overcome this challenge, our OCE system is operated at the shot noise limit and uses appropriate M-mode averaging to enhance sensitivity without adding motion artifacts. We also optimized the shape and dimensions of the actuator tip to maximize the excitation efficiency at high frequencies. For efficient generation of high frequency waves, the excitation stress profile must be well matched to the stress profile of the wave. This condition is achieved when the contact length is approximately a half wavelength. These advances enabled us to make OCE measurements of the skin at frequencies up to 10 kHz.

It is interesting to compare our results with previously reported values of skin tissue stiffness in the literature (see Appendix C Table C1). Our measured Young's moduli of the dermis and hypodermis are in reasonable agreement with previous results obtained at lower frequencies. However, our values for the epidermis (~ 3.5 MPa) are significantly higher than those obtained using lower frequency measurements. In particular, the previous OCE measurement in a 0.2–4 kHz range [17] estimated the epidermis stiffness to be ~ 200 kPa. We suspect that this rather low value is due to the limited frequency causing some averaging of epidermis and dermis, resulting in underestimation of the epidermis stiffness. Because of the viscoelasticity of skin tissues, the Young's modulus is expected to be significantly higher at 1 kHz than at 1 Hz [32]. Hence, our stiffness values are expected to be higher compared to quasi-static stiffness values. Indeed, compression-based OCE [33] measures Young's modulus at < 1 Hz.

Our results showed that the epidermis is about 100 times stiffer than the dermis. This is because of the contribution of the stiff stratum corneum (SC). A normal, dry SC has a considerably higher stiffness, between 100 MPa and 1 GPa [34]. Across most of the body, including the forearm, the thickness of the SC is ~ 20 µm [35], which is too thin to be resolved by OCE at frequencies up to 10 kHz. Therefore, our measured stiffness is an average of the SC and the viable epidermis. However, the SC thickness of palm, fingertip, and sole can be as large as 200 µm [2, 35]. In this case, the SC should be modeled as an individual layer, treating the viable epidermis as a separate layer. It would be interesting to extend our current 3-layer model algorithm to a 4-layer skin model including the SC.

Our current model assumes that the layers making up skin tissue are elastic, linear, and isotropic. Fitting the dispersion curve of a highly viscoelastic material with elastic theory will overestimate its storage modulus. The current dual bilayer model can be improved by taking into account the representative viscosity values of skin layers [36, 37].

Due to the limited imaging depth of our OCT system, the thickness of the dermis might have been underestimated. If we introduce a +10% variation for the dermal thickness (from 1 to 1.1 mm), the estimated Young's modulus would be 4.69 MPa (+3%) for the epidermis and 48 kPa (-6%) for the dermis for Subject #1; and 3.62 MPa for the epidermis (+2%) and 26 kPa (-7%) for the dermis for Subject #2. The overestimation of the dermal thickness results in overestimating the stiffness of the epidermis and underestimating the stiffness of the dermis.

While we have excluded the fast waves ('F' waves in Fig. 7) in this study, the supershear leaky surface waves can provide additional mechanical information about the skin. Recently we discovered that these waves are highly sensitive to the mechanical anisotropy and local stress [30]. The in-plane strain of forearm skin is less than 5% [38]. At this low strain level, the influence of prestress on Young's modulus is deemed small [39]. We plan to use supershear surface waves to characterize the stress field in the skin in a future study.

The broadband surface-wave OCE technique introduced here are expected to have several applications. For instance, it may prove useful for evaluating the influence of age, gender, and body site on the mechanical properties of the skin and for studying the impact of skin hydration and dryness [40], which is of great interest in the cosmetics industry. Finally, besides dermatologic applications, this technique may be extended to characterizing other layered tissues, such as blood vessel walls, and other soft materials with a depth-dependent stiffness gradient.

## Appendix

### A. Surface waves in a semi-infinite bilayer structure

We consider surface waves within the $x - z$ plane, propagating along the $x$ axis. We introduce two potential functions $\varphi_1$ and $\varphi_2$, and two stream functions $\psi_1$ and $\psi_2$ to decouple the transverse (shear) and longitudinal waves, which relate to the displacements of the film and the substrate by

$$u_x^f = \frac{\partial \varphi_1}{\partial x} + \frac{\partial \psi_1}{\partial z}, \quad u_z^f = \frac{\partial \varphi_1}{\partial z} - \frac{\partial \psi_1}{\partial x}, \quad u_x^s = \frac{\partial \varphi_2}{\partial x} + \frac{\partial \psi_2}{\partial z}, \quad u_z^s = \frac{\partial \varphi_2}{\partial z} - \frac{\partial \psi_2}{\partial x} \tag{A.1}$$

where $\mathbf{u}^f$ and $\mathbf{u}^s$ denote the displacements in the film and the substrate, respectively. Then the equilibrium equations expressed in the forms of the potential and stream functions are

$$\frac{\partial^2 \varphi_i}{\partial x^2} + \frac{\partial^2 \varphi_i}{\partial z^2} = \frac{1}{v_{li}^2} \frac{\partial^2 \varphi_i}{\partial t^2}, \quad \frac{\partial^2 \psi_i}{\partial x^2} + \frac{\partial^2 \psi_i}{\partial z^2} = \frac{1}{v_{ti}^2} \frac{\partial^2 \psi_i}{\partial t^2} \tag{A.2}$$

where $v_{li} = \sqrt{(\lambda_i + 2\mu_i)/\rho_i}$, and $v_{ti} = \sqrt{\mu_i/\rho_i}$ are the speeds of the longitudinal and transverse waves ($i = 1, 2$). $\lambda_1, \mu_1$ and $\lambda_2, \mu_2$ are the *Lamé* constants of the film and the substrate, respectively. $\rho_i$ is the density. $t$ is the time. Notice $\mu_1$ and $\mu_2$ are related to the Young's modulus of the top layer $E_1$ and the substrate $E_2$ by

$$\mu_1 = \frac{E_1}{2(1+\nu_1)}, \quad \mu_2 = \frac{E_2}{2(1+\nu_2)} \tag{A.3}$$

where the Poisson's ratios $\nu_1$ and $\nu_2$ are close to 0.5 for soft tissues. Consider the plane waves propagating along the $x$ axis, i.e., $\varphi_i = \varphi_{0i}(z) \exp[j(kx - \omega t)]$ and $\psi_i = \psi_{0i}(z) \exp[j(kx - \omega t)]$ ($i = 1, 2$), where $\omega$ and $k$ denote the angular frequency and wavenumber. Frequency is $f = \omega/(2\pi)$. Phase velocity is $v = \omega/k$. $j = \sqrt{-1}$. $\varphi_{0i}$ and $\psi_{0i}$ denote the wave amplitudes, of which the explicit expressions can be obtained by solving Eq. (2)

$$\varphi_{01} = a_1 e^{-i\alpha_1 z} + b_1 e^{i\alpha_1 z}, \quad \psi_{01} = c_1 e^{-i\beta_1 z} + d_1 e^{i\beta_1 z},$$

$$\varphi_{02} = a_2 e^{-i\alpha_2 z} + b_2 e^{i\alpha_2 z}, \quad \psi_{02} = c_2 e^{-i\beta_2 z} + d_2 e^{i\beta_2 z}, \tag{A.4}$$

where $\alpha_i = \sqrt{(\omega/v_{li})^2 - k^2}$, $\beta_1 = \sqrt{(\omega/v_{t1})^2 - k^2}$, and $\beta_2 = \sqrt{(\omega/v)^2 - k^2}$. $b_2 = 0$ and $d_2 = 0$ for the semi-infinite configuration. At the interface of the two layers, the displacements and stresses must be continuous, which yields

$$u_x^f = u_x^s, \quad u_z^f = u_z^s, \quad \sigma_{zz}^f = \sigma_{zz}^s, \quad \sigma_{xz}^f = \sigma_{xz}^s, \tag{A.5}$$

and at the free surface, the following stress-free boundary conditions apply

$$\sigma_{zz}^f = 0, \quad \sigma_{xz}^f = 0. \tag{A.6}$$

The Cauchy stresses $\sigma_{xz}^i$ and $\sigma_{zz}^i$ ($i = 1, 2$) are determined via Hooke's law:

$$\sigma_{zz}^i = (\lambda + 2\mu)\frac{\partial u_z^i}{\partial z} + \lambda\frac{\partial u_x^i}{\partial x}, \quad \sigma_{xz}^i = 2\mu\left(\frac{\partial u_x^i}{\partial z} + \frac{\partial u_z^i}{\partial x}\right). \tag{A.7}$$

Taking Eqs. (A.1), (A.4 – A.7), we get the following linear equation

$$\mathbf{M}_{6\times 6}[a_1, b_1, c_1, d_1, a_2, c_2]^{\mathrm{T}} = 0, \tag{A.8}$$

where

$$\mathbf{M} = \begin{bmatrix} k & k & -\beta_1 & \beta_1 & -k & \beta_2 \\ -\alpha_1 & \alpha_1 & -k & -k & \alpha_2 & k \\ \mu_1(k^2 - \beta_1^2) & \mu_1(k^2 - \beta_1^2) & -2\mu_1 k\beta_1 & 2\mu_1 k\beta_1 & -\mu_2(k^2 - \beta_2^2) & 2\mu_2 k\beta_2 \\ 2\mu_1 k\alpha_1 & -2\mu_1 k\alpha_1 & \mu_1(k^2 - \beta_1^2) & \mu_1(k^2 - \beta_1^2) & -2\mu_2 k\alpha_2 & -\mu_2(k^2 - \beta_2^2) \\ \mu_1(k^2 - \beta_1^2)e^{-i\alpha_1 h} & \mu_1(k^2 - \beta_1^2)e^{i\alpha_1 h} & -2\mu_1 k\beta_1 e^{-i\beta_1 h} & 2\mu_1 k\beta_1 e^{i\beta_1 h} & 0 & 0 \\ 2k\alpha_1 e^{-i\alpha_1 h} & -2k\alpha_1 e^{i\alpha_1 h} & (k^2 - \beta_1^2)e^{-i\beta_1 h} & (k^2 - \beta_1^2)e^{i\beta_1 h} & 0 & 0 \end{bmatrix}. \tag{A.9}$$

The existence of non-trivial solutions to Eq. (7) requires that the determinant of $\mathbf{M}$ is zero,

$$\det(\mathbf{M}) = 0, \tag{A.10}$$

From this equation, the dispersion relation of the surface waves is obtained.

**B. Equivalent thickness**

To obtain an equivalent thickness $\bar{h}_D$ for the dermis in the D-H bilayer, we take the long wavelength approximation. The bending stiffness of the E-D bilayer, denoted by $K$, is

$$K = \frac{1}{3}(E_D h_D^3 + E_E h_{DE}^3 - E_E h_D^3) - \frac{1}{4}\frac{(E_D h_D^2 + E_E h_{DE}^2 - E_E h_D^2)^2}{E_D h_D + E_E h_{DE} - E_E h_D}, \tag{B.1}$$

where $h_{DE} = h_D + h_E$. $K$ governs the flexural deformation, playing a dominant role in determining the surface wave speed. The bending stiffness of the equivalent dermis layer is $\frac{1}{12}E_D \bar{h}_D^3$, which should be equal to $K$. Therefore, we obtain the equivalent thickness

$$\bar{h}_D = \left(\frac{12K}{E_D}\right)^{1/3}. \tag{B.2}$$

It can be written as

$$\bar{h}_D = h_D \left(\frac{4(E_D + E_E(1+h_E/h_D)^3 - E_E)}{E_D} - \frac{3(E_D + E_E(1+h_E/h_D)^2 - E_E)^2}{E_D(E_D + E_E(1+h_E/h_D) - E_E)}\right)^{1/3}. \tag{B.3}$$

For $E_D = E_E$, we find $K = \frac{1}{12}E_D h_{DE}^3$ and $\bar{h}_D = h_D + h_E$.

For thin epidermis, a Taylor expansion up to the second order of $h_E/h_D$ gives

$$\bar{h}_D \approx h_D \left( \frac{1 + 4\frac{E_E h_E}{E_D h_D} + 6\frac{E_E h_E^2}{E_D h_D^2}}{1 + \frac{E_E h_E}{E_D h_D}} \right)^{1/3}. \tag{B.4}$$

For samples with very thin epidermis $h_E \ll h_D$ and $E_E h_E \gg E_D h_D$, we find $\bar{h}_D \approx 4^{1/3} h_D \approx 1.59\, h_D$. For thicker epidermis, yet thinner than dermis ($h_E < h_D$), we obtain $\bar{h}_D \approx 1.7\, h_D$.

## C. Previously reported stiffness values of human forearm skin in the literature

**Table C1.** Young's modulus of forearm skin in the literature

| Reference | # Subjects | Young's modulus (kPa) | Method | Frequency |
|---|---|---|---|---|
| Li [13] | 12 | 60 – 88 (dermis)<br>6 – 14 (hypodermis) | Shear wave elastography | ~ 1 kHz |
| Zhou [17] | 11 | 214 ± 106 (epidermis)<br>49 ± 26 (dermis)<br>10 ± 4 (hypodermis) | Surface acoustic wave elastography | < 4 kHz |
| Li [33] | 1 | ~ 500 (epidermis)<br>~ 150 (dermis)<br>~ 50 (hypodermis) | Compression OCE<br>Surface acoustic wave elastography | Quasi-static<br>< 4 kHz |
| Li [14] | 5 | 150 – 286 (dermis)<br>49 – 58 (hypodermis) | Surface acoustic wave elastography | < 4 kHz |
| Chartier [11] | 1 | ~ 142 (dermis)<br>~ 20 (hypodermis) | Transient elastography | 100 – 500 Hz |
| Liu [18] | 8 | 10 ± 4 (bulk) | Air-puff OCE | ~400 Hz |
| Liang [15] | 1 | 69 ± 25 (bulk) | Dynamic OCE | 50 Hz |
| Zhang [41] | 30 | 14 ± 10 (bulk) | Surface acoustic wave elastography | 100 – 400 Hz |
| Boyer [42] | 46 | 5 – 13 (bulk) | Dynamic indentation | 10 Hz |
| Hendriks [43] | 10 | ~ 56 (bulk) | Suction test | Quasi-static |
| Sanders [44] | 19 | 23 – 107 (bulk) | Torsion test | Quasi-static |

**Funding.** National Institutes of Health (NIH) (R01-EY025454, P41-EB015903).

**Disclosures.** The authors declare that there are no conflicts of interest related to this article.

**Supplemental document.** See **Supplement 1** for supporting content.

Supplementary Information

# *In vivo* stiffness measurement of epidermis, dermis, and hypodermis using broadband Rayleigh-wave optical coherence elastography

Xu Feng,[a,c,1] Guo-Yang Li,[a,c,1] Antoine Ramier,[a,b] Amira M. Eltony,[a,c] and Seok-Hyun Yun,[a,b,c,*]

[a] Harvard Medical School and Wellman Center for Photomedicine, Massachusetts General Hospital, Boston, Massachusetts 02114, USA

[b] Harvard-MIT Division of Health Sciences and Technology, Cambridge, Massachusetts 02139, USA

[c] Department of Dermatology, Massachusetts General Hospital, Boston, Massachusetts 02114, USA

[1] Co-first authors with equal contribution.

* syun@hms.harvard.edu

## S1. Finite element analysis

We built a plane strain model to study the wave propagation in the three-layer model. As shown in Fig. S1a, the width of the model is denoted by $w$. We scaled the size of the model using the wavelength $\lambda$, i.e., $w = 15\,\lambda$. The thickness of the third (hypodermis) layer $h_H = 0.5w$, which is large enough to avoid the reflections from the bottom. A time-harmonic pressure $p(t)$ with a Gaussian distribution (radius ~ 0.1 $\lambda$) was applied on the surface to excite the wave propagation. We used a symmetric boundary condition on the left side of the model. Other boundaries were stress free. The element type used in this study was 8-node biquadratic element (CPE8RH). Since only the surface wave motion was interested, we adopted a gradient mesh (Fig. S1b) to reduce the computation cost. The minimum size of the element was ~ 0.02 $\lambda$.

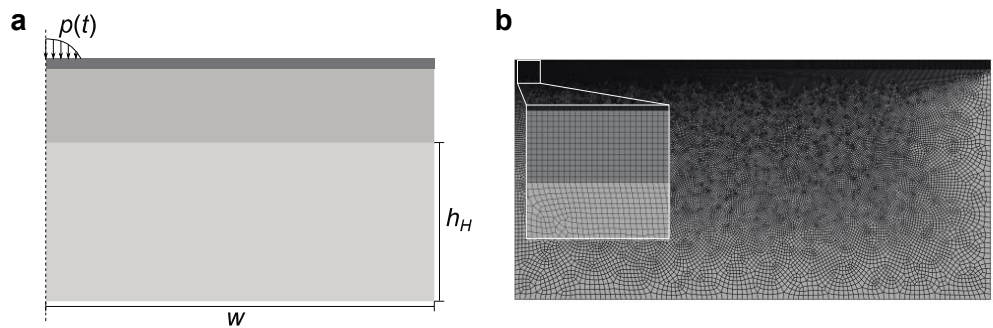

Fig. S1. Finite element analysis. (a) Boundary conditions and loads. (b) A representative mesh of the finite element model.

## S2. Mechanical characterization of the phantoms

Figure S2 displays the raw data used for estimating the reference Young's moduli of the individual phantom layers in Table 1. Figure S2 (a) shows the phase velocity dispersion curve of the PDMS film (thickness 150 $\mu m$) and the fitting curve with a Lamb wave model [1]. The estimated shear wave velocity was $19.7 \pm 0.7$ $m/s$. Figure S2 (b) shows the dispersion curve of the bulk gelatin phantoms (35 $mm$ in diameter and 10 $mm$ in height). The Rayleigh-type elastic wave speed $v_R$ was determined from the average wave speed in the frequency range of 1 to 10 kHz. The shear wave velocity is then determined as $v_S = v_R/0.955$ [2]. The measured shear wave velocity of the 7% and 3% bulk gelatin phantom were $3.0 \pm 0.1$ $m/s$ and $1.7 \pm 0.1$ $m/s$, respectively. The Young's modulus is then given by $E = 3\rho v_S^2$, where $\rho$ is the density ($\rho = 965$ $kg/m^3$ for PDMS and 1000 $kg/m^3$ for gelatin). We have determined that Young's modulus of the PDMS film is $1.13 \pm 0.08$ $MPa$, 7% gelatin phantom has $27 \pm 2$ $kPa$, and 3% gelatin phantom has $9 \pm 1$ $kPa$.

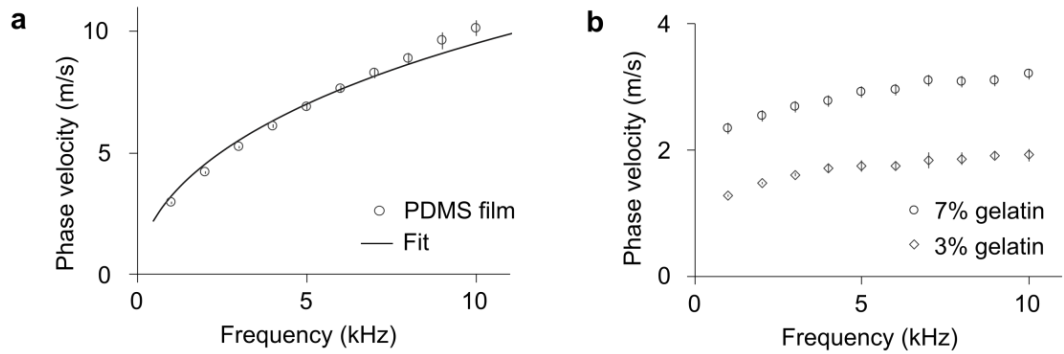

Fig. S2. Mechanical characterization of the materials used in the phantoms. (a) PDMS thin film. Measured dispersion curve (dot) is fitted by the Lamb wave model (solid line). (b) Measured phase velocity dispersion curve for the 7% gelatin phantom (dots) and the 3% gelatin phantom (diamonds). Error bars represent the standard deviation of three measurements at three different locations.

## S3. Comparison between finite element analysis and experimental data of human subjects

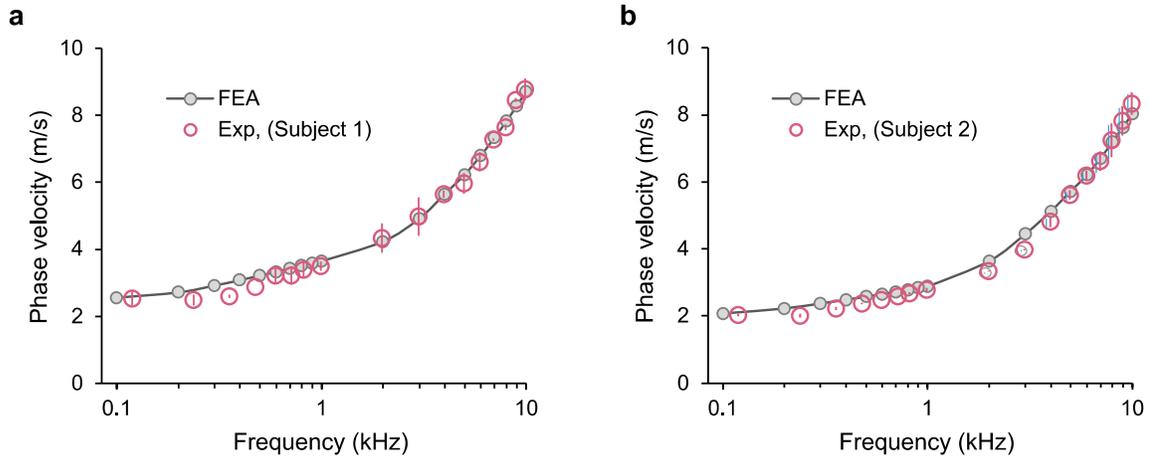

Fig. S3. Comparisons between the finite element analyses (FEA) and experimental data. (a) Subject #1 and (b) Subject #2. The FEA results were obtained using the fitting parameters in Table 3 in the main text.